\documentclass[11pt]{article}
\usepackage[latin9]{inputenc}
\usepackage{amsfonts}
\usepackage{amsmath}
\usepackage{amssymb}
\usepackage{indentfirst}
\usepackage{graphicx}
\usepackage[colorlinks]{hyperref}
\usepackage{cite}

\usepackage{cancel}

\numberwithin{equation}{section}
\usepackage{tikz}

\begin{document}

\begin{titlepage}
\vspace{3cm}
\baselineskip=24pt

\begin{center}
\textbf{\LARGE{Einstein manifolds with torsion and nonmetricity}}
\par\end{center}{\LARGE \par}

\begin{center}
	\vspace{1cm}
	\textbf{Dietmar Silke Klemm}$^{1,2}$,
	\textbf{Lucrezia Ravera}$^{2}$
	\small
	\\[5mm]
	$^{1}$\textit{Dipartimento di Fisica, Universit\`{a} di Milano, }\\
	\textit{ Via Celoria 16, 20133 Milano, Italy.}
	\\[2mm]
    $^{2}$\textit{INFN, Sezione di Milano, }\\
	\textit{ Via Celoria 16, I-20133 Milano, Italy.}
	\\[5mm]
	\footnotesize
	\texttt{dietmar.klemm@mi.infn.it},
	\texttt{lucrezia.ravera@mi.infn.it}
	\par\end{center}
\vskip 20pt
\begin{abstract}
\noindent

Manifolds endowed with torsion and nonmetricity are interesting both from the physical and the mathematical points of view. In this paper, we generalize some results presented in the literature. 
We study Einstein manifolds (i.e., manifolds whose symmetrized Ricci tensor is proportional to the metric) in $d$ dimensions with nonvanishing torsion that has both a trace and a 
traceless part, and analyze {invariance under extended conformal transformations} of the corresponding field equations.
Then, we compare our results to the case of Einstein manifolds with zero torsion and nonvanishing
nonmetricity, where the latter is given in terms of the Weyl vector (Einstein-Weyl spaces). We find that the
trace part of the torsion can alternatively be interpreted as the trace part of the nonmetricity.
The analysis is subsequently extended to Einstein spaces with both torsion and nonmetricity, where we also discuss the general setting in which the nonmetricity tensor has both a trace and a traceless part. Moreover, we consider and investigate actions involving scalar curvatures obtained from torsionful
or nonmetric connections, analyzing their relations with other gravitational theories that appeared previously in the literature. In particular, we show that the Einstein-Cartan action and the scale invariant gravity ({also known as} conformal gravity) action describe the same dynamics. Then, we consider the Einstein-Hilbert action coupled to a three-form field strength and show that its equations of motion imply that the manifold is Einstein with {totally antisymmetric} torsion.

\end{abstract}
\end{titlepage}\newpage {}

\section{Introduction}

In the $19$th century, the branches of mathematics and physics experienced an extraordinary progress with the emergence of non-Euclidean geometry. In particular, the development of Riemannian geometry
led to many important results, among which is the rigorous mathematical formulation of Einstein's general 
relativity.

In spite of the success and predictive power of general relativity, there are still some open problems and questions, whose understanding and solution may need the formulation of a new theoretical framework
as well as generalizations and extensions of Riemannian geometry. One possible way of generalizing 
Riemannian geometry consists in allowing for nonvanishing torsion and nonmetricity (metric affine
gravity) \cite{Hehl:1994ue} (see also \cite{A, B, C, D, Neeman:1996zcr, Sobreiro:2007cm, Puetzfeld:2014qba} and the recent work \cite{Jarv:2018bgs}).
There are several physical (and mathematical) reasons which motivate the introduction of torsion or
nonmetricity in the context of gravitational theories (see \cite{Hehl:1994ue} for details). For instance,
nonmetricity is a measure for the violation of local Lorentz invariance, which has been attracting some 
interest recently. Furthermore, nonmetricity and torsion find applications in the theory of defects in crystals, 
where, in particular, nonmetricity describes the density of point defects, while torsion is interpreted as 
density in line defects \cite{Kroner}. Moreover, as shown in \cite{Latorre:2017uve}, incorporating 
torsion and nonmetricity may allow to explore new physics associated with defects in a hypothetical 
spacetime microstructure. Recently, in \cite{Luz:2017ldh, Speziale:2018cvy, Iosifidis:2018diy} the authors discussed the propagation of matter fields in theories with torsion and nonmetricity. Further applications include quantum gravity \cite{Pagani:2015ema} and cosmology \cite{Stelmach:1991dx, Poberii:1994wp, BeltranJimenez:2017vop}. 

Moreover, torsion is related to the translation group and to the energy-momentum tensor of matter, while nonmetricity is related to the group $\rm{GL}(4,\mathbb{R})/ \rm{SO}(3,1)$ (in four dimensions) and to the hypermomentum current (see Refs. \cite{D} and \cite{Neeman:1996zcr, Puetzfeld:2014qba}, where, in particular in the latter, equations of motion in metric affine manifolds were studied); the trace of the nonmetricity (the Weyl vector) is related to the scale group and to the dilation (or scale) current. In particular, in matter the shear and dilation currents couple to nonmetricity, and they are its sources. It is to the dilation current that the Weyl vector is coupled.

Historically, a remarkable generalization of Riemannian geometry was first proposed in $1918$ by
Weyl (cf.~e.g.~\cite{Adler:1965, CP1, Folland, Romero:2012hs} for an introduction), who introduced an 
additional symmetry in an attempt of unifying electromagnetism with gravity
geometrically \cite{Weyl:1918ib, Weyl:1919fi}. In Weyl's theory, both the direction and the length of vectors 
are allowed to vary under parallel transport. However, Weyl's attempt to identify the trace part of the
connection associated with stretching and contraction with the vector potential of electromagnetism failed, due to observational inconsistencies (see e.g.~\cite{Wheeler:2018rjb} for details). Subsequently,
there were many attempts to adjust the theory. Finally, following \cite{London:1927fk}, Weyl showed that
a satisfactory theory of electromagnetism can be achieved if the scale factor is replaced by a complex phase. 
This was the origin of what is now well-known as the $\text{U} (1)$ gauge
theory\footnote{See \cite{ORaifeartaigh:1997dvq} and references therein for interesting details on `the dawning of gauge theory'.}.

The trace part of the connection introduced by Weyl is known as the Weyl vector. When it is given by the 
gradient of a function, there exists a scale transformation (dilatation) that sets the vector to zero. In this
case, Weyl geometry is said to be integrable (parallel transported vectors along closed paths return with unaltered lengths) and there exists a subclass of global gauges in which the geometry is Riemannian.

Although Weyl's theory of electromagnetism fails, there has been a renewed interest in
it \cite{Scholz:2017pfo, Barcelo:2017tes}. Indeed, there are motivations for seeking a deeper understanding 
of general relativity formulated within the framework of Weyl geometry (and especially of integrable Weyl geometry), in particular concerning scale invariant general relativity and higher symmetry approaches to 
gravity involving conformal invariance \cite{Wheeler:2018rjb}.
Always in Weyl's perspective, conformal (higher curvature) gravity theories were constructed and studied in detail in \cite{Tanhayi:2012nn,Dengiz:2012jb,Dengiz:2016eoo}. Furthermore, in \cite{Lobo:2018zrz} an observational constraint to the non-integrability of lengths in the original Weyl theory was placed for the first time.

A Weyl manifold is a conformal manifold equipped with a torsionless but nonmetric connection, called
Weyl connection, preserving the conformal structure.
Then, it is said to be Einstein-Weyl if the symmetric, trace-free part of the Ricci tensor of this connection vanishes (and the symmetric part of the Ricci tensor of the Weyl connection is proportional to the metric).
Thus, Einstein-Weyl manifolds represent the analog of Einstein spaces in Weyl geometry and are less
trivial than the latter, which have necessarily constant curvature in three dimensions.

Einstein-Weyl spaces were studied in \cite{TOD, PT1, PT2, PS1, PS2, PPS1, PPS2, PPS3, Madsen, Calderbank:1999ad, Calderbank:2000tk, Calderbank:2001bm, CP2}, and they are also relevant
in the context of (fake) supersymmetric supergravity solutions \cite{Meessen:2009ma,Gutowski:2009vb,Grover:2009ms,Gutowski:2011gc,Dunajski:2016rtx,
Randall:2017zlm}. Einstein-Weyl geometry is particularly rich in three 
dimensions \cite{TOD, PT1}, where it has an equivalent formulation in twistor theory \cite{Hitchin}, which 
provides a tool for constructing selfdual four-dimensional geometries.
Selfdual conformal four-manifolds play a central role in low-dimensional differential geometry, and a key tool 
in this context is provided by the so-called Jones-Tod correspondence \cite{Jones:1985pla}, in which the 
reduction of the self-duality equation by a conformal vector field is given by the Einstein-Weyl equation together with the linear equation for an abelian monopole (in other words, the Jones-Tod correspondence is
a correspondence between a self-dual space with symmetry and an Einstein-Weyl space with a monopole). 
Einstein-Weyl structures are also related to certain integrable systems, like the $\text{SU}(\infty)$ Toda field 
equations \cite{Ward:1990qt} or the dispersionless Kadomtsev-Petviashvili equation \cite{Dunajski:2000rf}.

On the other hand, as already mentioned, another generalization of Riemannian geometry is given by the 
introduction of a nonvanishing torsion, which is the case for the Einstein-Cartan
theory \cite{C1,C2,C3,C4,C5}, where the geometrical structure of the manifold is modified by allowing
for an antisymmetric part
of the affine connection (see also \cite{Cai:2015emx} for a recent review on torsional constructions and metric affine gauge theories). Cartan suggested that spacetime torsion is related to the intrinsic angular 
momentum, before the concept of spin was introduced.
Cartan's theory was then reinterpreted as a theory of gravitation with spin and
torsion \cite{S1,Sciama:1964wt,Kibble:1961ba}. Subsequently, the introduction of a non-vanishing torsion
has been widely analyzed in general relativity
and in the setting of teleparallel gravities \cite{Hehl:1976kj,Shapiro:1998zh,Shapiro:2001rz, Capozziello:2001mq, Watanabe:2004nt, Jensen, Golovnev:2018red}, as well as in other contexts. In particular,
in \cite{Scherk:1974mc,Saa:1993mi} the torsion tensor was related to the Kalb-Ramond
field \cite{Kalb:1974yc}.
Furthermore, the relation between torsion and conformal symmetry was studied by several authors, and
it turned out that torsion plays an important role in conformal invariance of the action and behaves like an 
effective gauge field \cite{Obukhov:1982zn, Maluf:1985fj}. Subsequently it was shown that in the
nonminimally coupled metric-scalar-torsion theory, for some special choice of the action, torsion acts as a 
compensating field and the full theory is conformally equivalent to general relativity at a classical
level \cite{German:1986uw, HelayelNeto:1999tm}. More recently, in \cite{Fabbri:2011vk} the
metric-torsional conformal curvature of four-dimensional spacetime was constructed, and in \cite{Chakrabarty:2018ybk} different types of torsion were investigated, together with their effect on the dynamics and conformal properties of fields. Conformal invariance 
was also analyzed in generalizations of Einstein-Cartan spaces including nonmetricity \cite{Smalley:1986tr, Dereli:2001tz,Drechsler:1993jy,Moon:2009zq}, and in \cite{Iosifidis:2018zwo} an exhaustive classification of metric affine theories according to their scale symmetries was presented {(see also \cite{Iosifidis:2019jgi})}.
Finally, in a cosmological context, it was proposed
in \cite{Cacciatori:2005wz,Ivanov:2016xjm} that a nonvanishing torsion can serve as an origin for dark energy.
Let us also mention, here, that	a generic theory (without matter) involving terms quadratic in torsion and nonmetricity will be classically equivalent at low energy to Einstein's theory, as discussed in \cite{Percacci:2009ij} and references therein.
From a mathematical point of view, Einstein manifolds with skew-symmetric torsion {(i.e., totally antisymmetric torsion)} were analyzed in \cite{Ferreira:2011zzb, Agricola}.

Motivated by the fact that nonmetric and torsionful connections are interesting both from the physical and the mathematical point of view, in this paper we generalize some results presented previously in the literature.
In particular, we study Einstein manifolds in $d$ dimensions with nonvanishing torsion that has both a
trace and a traceless part, and we analyze {invariance under extended conformal transformations (see Refs. \cite{German:1986uw,Smalley:1986tr}, where these transformations are defined for metric affine spaces)} in this context. Then, we compare our
results to the case of Einstein spaces with zero torsion and nonvanishing nonmetricity, where the latter is 
given in terms of the Weyl vector.
We find that the trace part of the torsion can alternatively be interpreted as the trace part of the
nonmetricity. Subsequently, we extend our analysis to the case of Einstein manifolds with both torsion and nonmetricity (Einstein-Cartan-Weyl spaces), where we allow for both a trace and a traceless part of
the nonmetricity tensor. Finally, we construct and investigate actions involving scalar curvatures obtained 
from torsionful or nonmetric connections, and analyze their relations with other gravitational theories known in the literature. In particular, we consider the Einstein-Cartan action and discuss its relationship with scale 
invariant gravity (also known as conformal gravity, {which is invariant under Weyl transformations}) \cite{Dirac:1938mt, Dirac:1973gk, Mannheim:1989jh, Aluri:2008ks, Chimento:2004it, Wetterich:2002wm, Gunzig:2000kk, Faraoni:2001tq, Kelleher:2003ss, Kelleher:2003pc}, showing that they describe the same dynamics.
Then, we study the Einstein-Hilbert action coupled to a three-form $H_{\mu\nu\rho}$ and shew that its equations of motion imply that the manifold is Einstein with skew-symmetric torsion. Furthermore, it turns out that the equations of motion of Einstein gravity coupled to a three-form may also be retrieved from a constrained action that contains the scalar curvature of a connection with torsion. Let us specify that in this work we will focus on the vacuum, without considering matter.

The remainder of this paper is organized as follows: In section \ref{ECmani}, we consider Einstein
spaces with torsion that has both a trace and a traceless part. In particular, we find the field equations
satisfied by an Einstein-Cartan space. Then, the {invariance under extended conformal (Weyl) transformations} of the latter is studied and the results are compared to the case of Einstein-Weyl manifolds, which have nonvanishing nonmetricity but zero torsion. In section \ref{ECWmani}, we extend the analysis to Einstein-Cartan-Weyl manifolds, and add thereby also a traceless part to the nonmetricity tensor. In section \ref{ECactionphi}, the {Weyl} invariant Einstein-Cartan action is studied and shown to be equivalent to scale invariant gravity (i.e., 
conformal gravity), which involves the presence of a scalar field $\phi$.
Subsequently, in section \ref{ECactionH} we consider the Einstein-Hilbert action coupled to a three-form,
and show that the resulting field equations imply that the space is Einstein with torsion, where the latter
is proportional to $H_{\mu\nu\rho}$. We conclude our work with some comments and possible future developments. {In the appendix we collect some technical details.}

\section{Einstein manifolds with torsion}\label{ECmani}

We first consider a $d$-dimensional Einstein manifold with metric $g_{\mu \nu}$ and nonvanishing
torsion (i.e., a so-called Einstein-Cartan manifold)\footnote{Our convention for the metric signature
is $(-,+,+,\cdots,+)$.}. The connection $\Gamma^\lambda_{\; \mu\nu}$ can be decomposed as
\begin{equation}\label{connection}
\Gamma^\lambda_{\; \mu \nu} = \tilde{\Gamma}^\lambda_{\; \mu \nu} + N^\lambda_{\; \mu \nu},
\end{equation}
where $\tilde{\Gamma}^\lambda_{\; \mu \nu}$ are the connection coefficients of the Levi-Civita
connection (i.e., the Christoffel symbols) and $N^\lambda_{\; \mu \nu}$ is called the distortion. 
{Here, the latter can be written as}\footnote{{As we will see in sec. 3, in the case of torsionful, nonmetric connections the distortion is generally defined as $N_{\lambda \mu \nu} = \frac{1}{2} \left( T_{\nu \lambda \mu} - T_{\lambda \nu \mu} - T_{\mu \nu \lambda} \right) + \frac{1}{2} \left(Q_{\lambda \mu \nu} + Q_{\lambda \nu \mu} - Q_{\mu \lambda \nu}  \right)$, where $Q_{\lambda \mu \nu}$ is the nonmetricity tensor (we will introduce and define it later). In the present section we first restrict ourselves to the case of vanishing nonmetricity, namely we consider a metric, torsionful connection. The nonmetric torsion-free case (where $N_{\lambda \mu \nu} = \frac{1}{2} \left(Q_{\lambda \mu \nu} + Q_{\lambda \nu \mu} - Q_{\mu \lambda \nu}  \right)$) will be discussed at the end of the current section when we will explore Einstein-Weyl spaces.}} 
\begin{equation}\label{distortion}
{N_{\lambda \mu \nu} = \frac{1}{2} \left( T_{\nu \lambda \mu} - T_{\lambda \nu \mu} - T_{\mu \nu \lambda} \right) ,}
\end{equation}
where $T^\lambda_{\; \mu \nu} = e_a^{\; \lambda} T^a_{\; \mu \nu}$ is the torsion\footnote{$e_a^{\;\lambda}$ denotes the inverse vielbein and early latin indices $a,b,\ldots$ refer
to the tangent space. {The torsion 2-form is defined as $T^a = d e^a + \omega^a_{\phantom{a}b} \wedge e^b$.}}, antisymmetric in the last two indices,
\begin{equation}\label{torsion}
T^\lambda_{\; \mu \nu} = \Gamma^{\lambda}_{\;\mu \nu} - \Gamma^{\lambda}_{\; \nu \mu} .
\end{equation}
Let us also introduce the contorsion (or 
contortion), antisymmetric in the first two indices,
\begin{equation}\label{contorsion}
K_{\nu \lambda \mu} = \frac{1}{2} \left( T_{\nu \lambda \mu} - T_{\lambda \nu \mu} - T_{\mu \nu \lambda} \right) .
\end{equation}
{Observe that the distortion \eqref{distortion} can then be written as}
\begin{equation}\label{distcont}
{N_{\lambda \mu \nu} = K_{\nu \lambda \mu} . }
\end{equation}
In \cite{Ferreira:2011zzb, Agricola}, Einstein manifolds with skew-symmetric torsion were analyzed.
Below, we shall consider a general decomposition of the torsion tensor, which can be decomposed in a
traceless and a trace part as
\begin{equation}\label{dector}
T^\lambda_{\; \mu \nu} = \breve{T}^\lambda _{\; \mu \nu} + \frac{1}{d-1} \left( \delta^\lambda_{\; \nu} T_ \mu - \delta^\lambda_{\; \mu} T_\nu \right)\,.
\end{equation}
In particular, we have $\breve{T}^\nu_{\; \mu \nu}=0$ and $T_\mu \equiv T^\nu_{\; \mu \nu}$.
Notice that $2 N ^\lambda_{\; [\mu \nu]} = T^\lambda_{\; \mu \nu}$.  
{The distortion \eqref{distcont} becomes then}
\begin{equation}
N_{\lambda \mu \nu} = \frac{1}{2} \left( T_{\nu \lambda \mu} - T_{\lambda \nu \mu} - T_{\mu \nu \lambda} \right)  = K_{\nu \lambda \mu} = \frac{1}{2} \left( \breve{T}_{\nu \lambda \mu} - \breve{T}_{\lambda \nu \mu} - \breve{T}_{\mu \nu \lambda} \right)+ \frac{1}{d-1} \left( g_{\mu \nu}T_\lambda - g_{\mu \lambda} T_\nu \right) ,
\end{equation}
and thus \eqref{connection} {reads}
\begin{equation}
\Gamma^\lambda_{\; \mu \nu} = \tilde{\Gamma}^\lambda_{\; \mu \nu} + \frac{1}{2} \left( \breve{T}^{\; \lambda}_{\nu\;\;  \mu} - \breve{T}^{\lambda}_{\; \nu \mu} - \breve{T}_{\mu \nu}^{\;\; \lambda} \right)+ \frac{1}{d-1} \left( g_{\mu \nu}T^\lambda - \delta_{\mu}^{\; \lambda} T_\nu \right) .
\end{equation}

{The explicit expression for the Riemann tensor $\tilde{R}^\lambda_{\;\rho\mu\nu}=\partial_\mu\Gamma^\lambda_{\;\nu\rho} - \partial_\nu \Gamma^\lambda_{\;\mu\rho} + \Gamma^\lambda_{\;\mu\sigma}\Gamma^\sigma_{\;\nu\rho} - \Gamma^\lambda_{\;\nu\sigma}\Gamma^\sigma_{\;\mu\rho}$ of the Einstein-Cartan connection $\Gamma^\lambda_{\; \mu\nu}$ is given in the appendix (see eq. \eqref{riemanntors}). There as well as in the following, $\nabla$ denotes the covariant derivative of the Levi-Civita connection.
The corresponding Ricci tensor $R_{\rho \nu} = R^\mu_{\; \rho \mu \nu}$ is given by \eqref{riccitens}.}
In particular, one gets
\begin{equation}\label{riccitensAS}
R_{[\rho\nu]} = \frac{d-2}{d-1} \nabla_{[\nu}T_{\rho]} - \frac{1}{2}\breve{T}_{\mu \sigma [\nu} \breve{T}_{\rho]}^{\;\; \mu \sigma} - \frac{1}{d-1} T^\mu \breve{T}_{[\rho \nu] \mu} + \frac{2-d}{2(d-1)} T^\mu \breve{T}_{\mu \nu \rho} + \frac{1}{2} \nabla_\mu \breve{T}^\mu_{\; \nu \rho} .
\end{equation}
Note that if we set the traceless part of the torsion to zero, $\breve{T}^\lambda_{\;\mu\nu}=0$, we are left with
\begin{equation}\label{riccitensASnew}
R_{[\rho \nu]} = \frac{d-2}{d-1} \nabla_{[\nu}T_{\rho]} = \frac{d-2}{d-1} \partial_{[\nu}  T_{\rho]} \equiv \frac{d-2}{2(d-1)} F_{\nu \rho} ,
\end{equation}
where
\begin{equation}\label{Fnurho}
F_{\nu \rho} \equiv \partial_\nu T_\rho - \partial_\rho T_\nu .
\end{equation}
In general, one has thus
\begin{equation}
R_{[\rho\nu]} = \frac{d-2}{2(d-1)} F_{\nu \rho} - \frac{1}{2}\breve{T}_{\mu \sigma [\nu} \breve{T}_{\rho]}^{\;\; \mu \sigma} - \frac{1}{d-1} T^\mu \breve{T}_{[\rho \nu] \mu} + \frac{2-d}{2(d-1)} T^\mu \breve{T}_{\mu \nu \rho} + \frac{1}{2} \nabla_\mu \breve{T}^\mu_{\; \nu \rho} .
\end{equation}
One can also construct another Ricci tensor by contracting the second and the third index of the Riemann 
tensor. However, the Ricci tensor obtained in this way coincides with \eqref{riccitens}, since
$R_{\lambda\rho\mu\nu} = - R_{\rho\lambda\mu\nu}$ is still valid (while it fails to be for nonmetric
connections).

The Ricci scalar reads
\begin{equation}\label{ricciscalar}
R = g^{\rho \nu} R_{\rho \nu} = \tilde{R}  + \frac{(d-2)(1-d)}{(d-1)^2}T_\mu T^\mu + 2 \nabla_\mu T^\mu + \frac{1}{4} \breve{T}_{\mu \nu \rho} \breve{T}^{\mu \nu \rho} - \frac{1}{2} \breve{T}_{\nu \rho \mu} \breve{T}^{\mu \nu \rho} .
\end{equation}

Let us now define an Einstein space with torsion by
\begin{equation}\label{symmcond}
R_{(\rho \nu)} = \lambda g_{\rho \nu}
\end{equation}
for some function $\lambda$. Using \eqref{riccitens}, this becomes
\begin{equation}\label{Riccisymm}
\begin{split}
& \tilde{R}_{\rho \nu}  + \frac{1}{d-1} \left[ g_{\rho \nu} \nabla_\mu T^\mu + (d-2) \nabla_{(\nu} T_{\rho)} + (d-3) T^\mu \breve{T}_{(\rho \nu ) \mu} \right]  \\
& + \frac{1}{(d-1)^2} \left[ (2-d) g_{\rho \nu} T_\mu T^\mu + (d-2) T_{\nu} T_{\rho} \right] \\
& + \frac{1}{4} \breve{T}_{\rho}^{\;\; \mu \sigma} \breve{T}_{\nu \mu \sigma} - \frac{1}{2} \breve{T}_{\mu \sigma (\rho} \breve{T}_{\nu)}^{\;\;\;\mu \sigma} - \nabla_\mu \breve{T}_{(\rho \nu)}^{\;\;\;\;\;\mu}    = \lambda g_{\rho \nu} ,
\end{split}
\end{equation}
whose trace yields
\begin{equation}
\tilde{R} + 2\nabla_\mu T^\mu - \frac{d-2}{d-1} T_\mu T^\mu + \frac{1}{4} \breve{T}^{\mu \rho \sigma} \breve{T}_{\mu \rho \sigma} - \frac{1}{2} \breve{T}^{\mu \rho \sigma} \breve{T}_{\rho \sigma \mu} = \lambda d ,
\end{equation}
and thus
\begin{equation}\label{lambda}
\lambda = \frac1d\left( \tilde{R} + 2 \nabla_\mu T^\mu  - \frac{d-2}{d-1} T_\mu T^\mu    + \frac{1}{4} \breve{T}^{\mu \rho \sigma} \breve{T}_{\mu \rho \sigma} - \frac{1}{2} \breve{T}^{\mu \rho \sigma} \breve{T}_{\rho \sigma \mu} \right) .
\end{equation}
Hence, in terms of Riemannian data, \eqref{symmcond} becomes
\begin{equation}\label{pdeEC}
\begin{split}
& \tilde{R}_{\rho \nu}  + \frac{1}{d-1} \left[ (d-2) \nabla_{(\nu} T_{\rho)} + (d-3) T^\mu \breve{T}_{(\rho \nu ) \mu} \right]  + \frac{1}{(d-1)^2} \left[  (d-2) T_{\nu} T_{\rho} \right] \\
& + \frac{1}{4} \breve{T}_{\rho}^{\;\; \mu \sigma} \breve{T}_{\nu \mu \sigma}  - \frac{1}{2} \breve{T}_{\mu \sigma (\rho} \breve{T}_{\nu)}^{\;\;\;\mu \sigma} - \nabla_\mu \breve{T}_{(\rho \nu)}^{\;\;\;\;\;\mu}  \\
&  = \frac{1}{d} g_{\rho \nu}  \left[ \tilde{R} + \frac{d-2}{d-1} \nabla_\mu T^\mu  + \frac{d-2}{(d-1)^2} T_\mu T^\mu    + \frac{1}{4} \breve{T}^{\mu \tau \sigma} \breve{T}_{\mu \tau \sigma} - \frac{1}{2} \breve{T}^{\mu \tau \sigma} \breve{T}_{\tau \sigma \mu} \right] ,
\end{split}
\end{equation}
which is a set of nonlinear partial differential equations characterizing an Einstein manifold with
torsion, henceforth termed Einstein-Cartan space.

\subsection{Extended conformal invariance in Einstein-Cartan manifolds}

We will now show that \eqref{symmcond} is {invariant under extended conformal transformations discussed in \cite{German:1986uw}. Thus, let us consider the extended conformal (Weyl) transformations}
\begin{equation}\label{transf}
\begin{split}
& g_{\mu \nu}\mapsto g'_{\mu \nu}= e^{2 \omega} g_{\mu \nu} , \\
& T^\lambda_{\; \mu \nu}\mapsto T'^\lambda_{\; \mu \nu}  =  T^\lambda_{\; \mu \nu} + 
\delta^\lambda_{\; \nu} \partial_\mu \omega - \delta^\lambda_{\; \mu} \partial_\nu \omega ,
\end{split}
\end{equation}
{where $\omega = \omega(x)$ is an arbitrary scalar field. Therefore, we have}
\begin{equation}\label{transfTtildeT}
T_\mu\mapsto T'_\mu = T_\mu + (d-1) \partial_\mu \omega , \qquad
\breve{T}^\lambda_{\;\mu\nu}\mapsto\breve{T}'^\lambda_{\;\mu\nu} = \breve{T}^\lambda_{\;\mu\nu} .
\end{equation}
{Moreover, \eqref{transf} leads to the following transformation for the connection:}
\begin{equation}\label{spprtr}
\Gamma^{\rho}_{\; \mu \nu}\mapsto\Gamma'^{\rho}_{\; \mu \nu} = \Gamma^{\rho}_{\; \mu \nu} + \delta^\rho_{\; \nu} \partial_\mu \omega ,
\end{equation}
{which is called, specifically, a special projective transformation of the connection (see, for instance, Refs. \cite{Iosifidis:2018zwo, Iosifidis:2019jgi}), also known as $\lambda$ transformation.}
{Let us observe that, actually, the combination of the conformal metric transformation in \eqref{transf} plus the special projective transformation \eqref{spprtr} of the affine connection is called a frame rescaling (see Refs. \cite{Iosifidis:2018zwo, Iosifidis:2019jgi}, where frame rescalings have been considered in metric affine spaces, also including Einstein-Cartan ones).}

For the Riemann tensor, the Ricci tensor and the scalar curvature, we get respectively
\begin{equation}\label{transfR}
\begin{split}
& R^\sigma_{\;\rho\mu\nu}\mapsto R'^\sigma_{\;\rho\mu\nu} = R^\sigma_{\;\rho\mu\nu} , \\
& R_{\rho\nu}\mapsto R'_{\rho\nu} = R_{\rho\nu} , \\
& R\mapsto R' = e^{-2 \omega} R .
\end{split}
\end{equation}
Now, \eqref{symmcond} implies $R= \lambda d$, so that \eqref{symmcond} is equivalent to 
\begin{equation}
R_{(\rho \nu)} = \frac1d R g_{\rho \nu} ,
\end{equation}
which is obviously {invariant under extended conformal transformations given by \eqref{transf}.}

\subsection{Comparison with Einstein-Weyl spaces}
\label{subsec:compEW}

A Weyl structure on a manifold $\Sigma$ consists of a conformal structure
$[g]=\{fg| f:\Sigma\to\mathbb{R}^+\}$, and a torsion-free connection $\hat\nabla$ {fulfilling}
\begin{equation}\label{weylconn}
{\hat{\nabla}_\nu g_{\lambda \mu} = 2 \Theta_\nu g_{\lambda \mu} , }
\end{equation}
{for some one-form $\Theta$ on $\Sigma$ (the Weyl vector). The condition \eqref{weylconn} is invariant under the transformation}
\begin{equation}
g_{\mu\nu}\mapsto g'_{\mu\nu} = e^{2\omega} g_{\mu\nu} , \qquad
\Theta_\mu\mapsto\Theta'_\mu = \Theta_\mu + \partial_\mu\omega .
\end{equation} 
{One can then define the nonmetricity tensor, which reads}
\begin{equation}\label{nonmet}
Q_{\mu \nu \lambda} = - \hat{\nabla}_\nu g_{\lambda \mu} = - 2 \Theta_\nu g_{\lambda \mu} . 
\end{equation}
{In this case the distortion is given by}
\begin{equation}\label{distortionEW}
{N^\lambda_{\; \mu \nu} = \frac{1}{2} \left(Q_{\lambda \mu \nu} + Q_{\lambda \nu \mu} - Q_{\mu \lambda \nu}\right) = - \delta^{\lambda}_{\;\nu} \Theta_\mu - \delta^\lambda_{\;\mu} \Theta_\nu + \Theta^\lambda g_{\mu\nu}.}
\end{equation}

A Weyl structure is said to be Einstein-Weyl \cite{CP1} if the symmetrized Ricci tensor $W_{\rho\nu}$ of 
$\hat\nabla$ is proportional to some metric $g\in[g]$,
\begin{equation}\label{symmcondEW}
W_{(\rho\nu)} = \frac1d g_{\rho\nu} W , 
\end{equation}
where $W$ is the scalar curvature of the Weyl connection $\hat\nabla$. It is given by\footnote{See also the 
results of sec.~\ref{ECWmani} in the case of zero torsion.}
\begin{equation}\label{WRicci}
W = \tilde{R} + (d-2)(1-d)\Theta_\mu\Theta^\mu + 2(d-1)\nabla_\mu\Theta^\mu .
\end{equation}
{The condition} \eqref{symmcondEW} can be rewritten in terms of Riemannian
data as
\begin{equation}\label{pdeEW}
\tilde{R}_{\rho\nu} + (d-2)\Theta_{\rho}\Theta_{\nu} + (d-2)\nabla_{(\nu}\Theta_{\rho)} = \frac1d
g_{\rho\nu}\left[\tilde{R} + (d-2)\nabla_\mu\Theta^\mu + (d-2)\Theta_\mu\Theta^\mu\right] .
\end{equation}

The scope of this subsection is to compare the field equations for Einstein manifolds with torsion,
\eqref{pdeEC}, with the Einstein-Weyl equations \eqref{pdeEW}. To this end, let us define
\begin{equation}\label{defA}
A_\mu \equiv \frac{T_\mu}{d-1},
\end{equation}
such that, under the {first transformation in \eqref{transfTtildeT}}, we have
\begin{equation}\label{transfA}
A_\mu\mapsto A'_\mu = A_\mu + \partial_\mu \omega .
\end{equation}
Using \eqref{defA} in \eqref{pdeEC}, one gets
\begin{equation}\label{pdeECnew}
\begin{split}
& \tilde{R}_{\rho\nu} + (d-2) A_{\rho} A_{\nu} + (d-2)\nabla_{(\nu} A_{\rho)} + (d-3) A^\mu
\breve{T}_{(\rho\nu)\mu} + \frac14\breve{T}_{(\rho}^{\;\;\;\mu\sigma}\breve{T}_{\nu)\mu\sigma} -
\frac12\breve{T}_{\mu\sigma(\rho}\breve{T}_{\nu)}^{\;\;\;\mu\sigma} - \nabla_\mu
\breve{T}_{(\rho\nu)}^{\;\;\;\;\;\mu}  \\
&  = \frac1d g_{\rho\nu}\left[\tilde{R} + (d-2)\nabla_\mu A^\mu + (d-2) A_\mu A^\mu + \frac14 
\breve{T}^{\mu\tau\sigma}\breve{T}_{\mu\tau\sigma} - \frac12\breve{T}^{\mu\tau\sigma}
\breve{T}_{\tau\sigma\mu} \right] .
\end{split}
\end{equation}
Thus, for $\breve{T}^\lambda_{\;\mu\nu}=0$, \eqref{pdeECnew} exactly coincides with \eqref{pdeEW} if
we identify $A_\mu$ with $\Theta_\mu$, i.e., $T_\mu\rightarrow (d-1)\Theta_\mu$.
This is actually not surprising, since for $\breve{T}^\lambda_{\;\mu\nu}=0$ the torsion two-form
is given by
\begin{equation}
T^a_{\;\mu\nu} = \frac1{d-1}\left(e^a_{\;\nu} T_\mu - e^a_{\;\mu} T_\nu\right) = e^a_{\;\nu} A_\mu - 
e^a_{\;\mu} A_\nu = -\left(e^a\wedge A\right)_{\mu\nu}\quad\Rightarrow\quad T^a = A\wedge e^a .
\end{equation}
Then, the first Cartan structure equation gives
\begin{equation}
d e^a + {\omega^a}_b\wedge e^b = A\wedge e^a \quad\Rightarrow\quad d e^a + \left({\omega^a}_b - 
{\delta^a}_b A\right)\wedge e^b =0 .
\end{equation}
We can then define a new connection $\hat{\omega}^{ab}$ as
\begin{equation}
\hat{\omega}^{ab} = \omega^{ab} - \eta^{ab} A ,
\end{equation}
which is torsion-free
\begin{equation}
d e^a + \hat{\omega}^a_{\,\,\,b}\wedge e^b =0 ,
\end{equation}
but nonmetric, since $\hat{\omega}^{(ab)}\neq 0$. The trace part of the torsion can thus always be
shuffled into a Weyl vector and vice-versa. In the latter case, a Weyl structure gets translated into a
conformal structure $[g]$ together with a torsionful connection $D$ which is compatible
with $[g]$,
\begin{equation}\label{Dg=0}
D_\mu g_{\nu \lambda} = 0.
\end{equation}
The torsion of $D$ has only a trace part $T_\mu$, and \eqref{Dg=0} is invariant under the {transformation} \eqref{transf}, \eqref{transfTtildeT}.

Finally, note that a duality between torsion and nonmetricity has also been discussed in \cite{Iosifidis:2018zjj} in a slightly different context.

\section{Einstein manifolds with torsion and nonmetricity}\label{ECWmani}

Let us now consider Einstein spaces with both torsion and nonmetricity (we will call these
Einstein-Cartan-Weyl manifolds), and study the {Weyl} invariance of the corresponding field equations.

With respect to section \ref{ECmani}, we will in addition allow for a nonmetricity tensor of the form
\eqref{nonmet}, where $\hat\nabla$ has also torsion. We are thus considering only the trace part
of the nonmetricity. The consequences of adding a traceless part will be analyzed at the end of this section.
The connection $\hat{\Gamma}^\lambda_{\;\mu\nu}$ of the Einstein-Cartan-Weyl manifold is given by
\begin{equation}\label{connectionECW}
\hat{\Gamma}^\lambda_{\;\mu\nu} = \tilde{\Gamma}^\lambda_{\;\mu\nu} + N^\lambda_{\;\mu\nu},
\end{equation}
where the $\tilde{\Gamma}^\lambda_{\;\mu\nu}$ are the Christoffel symbols, and the distortion
$N^\lambda_{\;\mu\nu}$ {reads}
\begin{equation}\label{distortioncompl}
{N_{\lambda \mu \nu} = \frac{1}{2} \left( T_{\nu \lambda \mu} - T_{\lambda \nu \mu} - T_{\mu \nu \lambda} \right) + \frac{1}{2} \left(Q_{\lambda \mu \nu} + Q_{\lambda \nu \mu} - Q_{\mu \lambda \nu}  \right) , }
\end{equation}
{that is, in the present context,}
\begin{equation}\label{distcontECW}
N_{\lambda\mu\nu} = \frac12\left(\breve{T}_{\nu\lambda\mu} - \breve{T}_{\lambda\nu\mu} - 
\breve{T}_{\mu\nu\lambda}\right) + \frac1{d-1}\left(g_{\mu\nu} T_\lambda - g_{\mu\lambda} T_\nu\right) 
+ \Theta_\lambda g_{\mu\nu} - \Theta_\mu g_{\lambda\nu} - \Theta_\nu g_{\lambda\mu} .
\end{equation}

{The Ricci tensor of $\hat\nabla$, that is $\hat{R}_{\rho \nu} = \hat{R}^\mu_{\; \rho \mu \nu}$, is given in the appendix (see eq. \eqref{riccitensECW}).}
Note that one can {also} construct another Ricci
tensor $\mathfrak{R}_{\rho\nu}=\hat{R}^\mu_{\;\mu\rho\nu}$ (commonly referred to as the homothetic 
curvature), since for nonmetric connections the Riemann tensor is not necessarily antisymmetric in
the first two indices. In our case we have
\begin{equation}\label{homot}
\mathfrak{R}_{\rho\nu} = d\left(\nabla_\nu\Theta_\rho - \nabla_\rho\Theta_\nu\right),
\end{equation}
and thus the Ricci scalar associated with the homothetic curvature is identically zero.
On the other hand, the nonvanishing Ricci scalar is given by
\begin{equation}\label{ricciscalarECW}
\begin{split}
\hat{R} & = g^{\rho\nu}\hat{R}_{\rho\nu} \\
& =\tilde{R} + \frac{(d-2)(1-d)}{(d-1)^2} T_\mu T^\mu + 2\nabla_\mu T^\mu + \frac14
\breve{T}_{\mu\nu\rho}\breve{T}^{\mu\nu\rho} - \frac12\breve{T}_{\nu\rho\mu}\breve{T}^{\mu\nu\rho} \\
& + (d-2)(1-d)\Theta_\mu\Theta^\mu + 2 (d-1)\nabla_\mu\Theta^\mu + 2(2-d)\Theta^\mu T_\mu .
\end{split}
\end{equation}
Observe that, if we define
\begin{equation}\label{tcheck}
\check{T}_\mu\equiv T_\mu + (d-1)\Theta_\mu ,
\end{equation}
the Ricci scalar \eqref{ricciscalarECW} becomes
\begin{equation}\label{tcheckeq}
\hat{R} = \tilde{R} + \frac{(d-2)(1-d)}{(d-1)^2}\check{T}_\mu\check{T}^\mu + 2\nabla_\mu 
\check{T}^\mu + \frac14\breve{T}_{\mu\nu\rho}\breve{T}^{\mu\nu\rho} - \frac12\breve{T}_{\nu\rho\mu} 
\breve{T}^{\mu\nu\rho} ,
\end{equation}
which corresponds to the Ricci scalar of a metric connection with torsion (cf.~eq.~\eqref{ricciscalar}), whose 
trace part is given by $\check{T}_\mu$.

We define an Einstein-Cartan-Weyl space by
\begin{equation}\label{symmcondECW}
\hat{R}_{(\rho \nu)} = \lambda g_{\rho \nu}
\end{equation}
for some function $\lambda$. Using \eqref{riccitensECW}, this can be rewritten in the equivalent form
\begin{equation}\label{pdeECW}
\begin{split}
& \tilde{R}_{\rho \nu}  + \frac{1}{d-1} \left[ (d-2) \nabla_{(\nu} T_{\rho)} + (d-3) T^\mu \breve{T}_{(\rho \nu ) \mu} \right]  + \frac{1}{(d-1)^2} \left[  (d-2) T_{\nu} T_{\rho} \right] \\
& + \frac{1}{4} \breve{T}_{\rho}^{\;\; \mu \sigma} \breve{T}_{\nu \mu \sigma} - \frac{1}{2} \breve{T}_{\mu \sigma (\rho} \breve{T}_{\nu)}^{\;\;\;\mu \sigma}  - \nabla_\mu \breve{T}_{(\rho \nu)}^{\;\;\;\;\;\mu}  \\
& + (d-2) \Theta_{\nu} \Theta_{\rho} + (d-2) \nabla_{(\nu} \Theta_{\rho)} + \frac{2(d-2)}{d-1} \Theta_{(\nu} T_{\rho)} + (d-3) \Theta^\mu \breve{T}_{(\nu \rho) \mu} \\
&  = \frac{1}{d} g_{\rho \nu}  \Bigg [ \tilde{R} + \frac{d-2}{d-1} \nabla_\mu T^\mu  + \frac{d-2}{(d-1)^2} T_\mu T^\mu    + \frac{1}{4} \breve{T}^{\mu \tau \sigma} \breve{T}_{\mu \tau \sigma} - \frac{1}{2} \breve{T}^{\mu \tau \sigma} \breve{T}_{\tau \sigma \mu} \\
& + (d-2)\nabla_\mu \Theta^\mu + (d-2) \Theta_\mu \Theta^\mu + \frac{2(d-2)}{d-1} \Theta^\mu T_\mu \Bigg ] ,
\end{split}
\end{equation}
which is a system of nonlinear partial differential equations characterizing an Einstein-Cartan-Weyl manifold.

\subsection{Extended conformal invariance of the Einstein-Cartan-Weyl equations}

Let us now discuss the {extended conformal} invariance of \eqref{symmcondECW}.
{In an affine manifold such as an Einstein-Cartan-Weyl one, the most general extended conformal (Weyl) transformations involving an arbitrary scalar field $\omega = \omega(x)$ which leave the curvature tensor invariant are given by (see \cite{Smalley:1986tr})}
\begin{equation}\label{transfECW}
\begin{split}
& g_{\mu \nu}\mapsto e^{2 \omega} g_{\mu \nu} , \\
& T^\lambda_{\;\mu\nu}\mapsto T^\lambda_{\;\mu\nu} + 2(1-\xi)\delta^\lambda_{\;[\nu}\partial_{\mu]} \omega , \\
& Q^\lambda_{\;\mu\nu} \mapsto Q^\lambda_{\;\mu\nu} - 2\xi\partial_\mu\omega\delta^\lambda_{\;\nu} ,
\end{split}
\end{equation}
{where $\xi$ denotes an arbitrary parameter that we are free to include \cite{Smalley:1986tr,Moon:2009zq}\footnote{{Note that \eqref{transfECW} implies that
$\hat\nabla_\mu g_{\nu\rho}=2\Theta_\mu g_{\nu\rho}$ transforms covariantly.}}.
In particular, for the one-forms $\Theta$ and $T$ and for $\breve{T}^\lambda_{\;\mu\nu}$ we find}
\begin{equation}\label{transfECWThTTbr}
\begin{split}
& \Theta_\mu\mapsto \Theta_\mu + \xi\partial_\mu\omega , \\
& T_\mu\mapsto  T_\mu + (1-\xi)(d-1)\partial_\mu\omega , \\
& \breve{T}^\lambda_{\;\mu\nu}\mapsto \breve{T}^\lambda_{\;\mu\nu} ,
\end{split}
\end{equation}
{and the connection $\hat\Gamma$ transforms} according to
\begin{equation}\label{transfECWconn}
\hat{\Gamma}^{\rho}_{\;\mu\nu} \mapsto \hat{\Gamma}^{\rho}_{\;\mu\nu} + (1-\xi)\delta^\rho_{\;\nu} 
\partial_\mu\omega .
\end{equation}
{This ensures the invariance of the curvature tensor due to its special projective invariance (see, for instance, Refs. \cite{Iosifidis:2018zwo, Iosifidis:2019jgi}).}

{Thus, for} the Riemann tensor, the Ricci tensor and the scalar curvature one obtains respectively
\begin{equation}\label{transfRECW}
\hat{R}^\sigma_{\;\rho\mu\nu}\mapsto \hat{R}^\sigma_{\;\rho\mu\nu} , \qquad
\hat{R}_{\rho\nu}\mapsto \hat{R}_{\rho\nu} , \qquad
\hat{R}\mapsto e^{-2\omega}\hat{R} .
\end{equation}
{Eq.} \eqref{symmcondECW} implies $\hat{R}=\lambda d$, so that \eqref{symmcondECW} is equivalent to 
\begin{equation}
\hat{R}_{(\rho\nu)} = \frac1d\hat{R} g_{\rho\nu} ,
\end{equation}
which is clearly {invariant under the extended conformal transformations written above}.

Let us finally make some comments on two particular cases, namely $\xi=1$ and $\xi=0$.
\begin{itemize}
\item For $\xi=1$ one has
\begin{equation}
T_\mu\mapsto T_\mu, \qquad
\Theta_\mu\mapsto\Theta_\mu + \partial_\mu\omega . \label{xi1T}
\end{equation}
Observe that \eqref{xi1T} corresponds to the transformation {\eqref{transfA}, for $A_\mu=\Theta_\mu$,} discussed in  section \ref{ECmani} in the context of a Weyl structure (that is with nonmetricity and zero torsion).
Moreover, note that this is the only case in which the connection is also invariant,
$\hat{\Gamma}^{\rho}_{\;\mu\nu}\mapsto\hat{\Gamma}^{\rho}_{\;\mu\nu}$. {In fact, setting $\xi=1$ into \eqref{transfECW} and \eqref{transfECWThTTbr} leads to a conformal transformation of the metric in an affine space, namely a transformation under which the metric tensor picks up a conformal factor $e^{2 \omega}$ while the affine connection is left unchanged (see Refs. \cite{Iosifidis:2018zwo, Iosifidis:2019jgi}).}
\item For $\xi=0$ we get {the extended conformal transformation discussed in \cite{German:1986uw} in the context of a torsion theory which leads to a special projective transformation for the connection. In particular, in this case we have}
\begin{equation}
T_\mu\mapsto T_\mu + (d-1)\partial_\mu\omega , \qquad
\Theta_\mu\mapsto\Theta_\mu ,
\end{equation} 
{which} reproduces exactly the transformation {in \eqref{transfTtildeT}} for $T_\mu$ discussed in section \ref{ECmani} for manifolds with torsion and vanishing nonmetricity, {together with}
\begin{equation}\label{xi0conn}
\hat{\Gamma}^{\rho}_{\;\mu\nu}\mapsto\hat{\Gamma}^{\rho}_{\;\mu\nu} + \delta^\rho_{\;\nu} 
\partial_\mu\omega {,}
\end{equation}
which is a special projective transformation \eqref{xi0conn} for the connection. On the other hand, let us observe that the combination of the conformal metric transformation in \eqref{transfECW} plus the special projective transformation \eqref{xi0conn} is called, according to \cite{Iosifidis:2018zwo, Iosifidis:2019jgi}, a frame rescaling.
\end{itemize}
{We can conclude that there are two unique transformations which single out torsion or nonmetricity.}
This is in agreement with \cite{Smalley:1986tr}.
{Note that the same results could have been obtained by considering 
\eqref{tcheckeq}, together with the definition \eqref{tcheck}, that is by reabsorbing the nonmetricity and 
exploiting the transformations of sec. \ref{ECmani} for an Einstein-Cartan manifold with torsion 
and vanishing nonmetricity.}

\subsection{Adding a traceless part to the nonmetricity tensor}

In the following we extend the above analysis to include a traceless part of the nonmetricity as well.
Interestingly, in the case where the latter is totally symmetric, it can be viewed as representing a massless 
spin-$3$ field \cite{Boulanger:2006tg, Baekler:2006vw}.

Thus, we decompose
\begin{equation}\label{nonmetricityECWfull}
Q_{\lambda\mu\nu} = -2\Theta_\mu g_{\nu\lambda} + \breve{Q}_{\lambda\mu\nu},
\end{equation}
where $\breve{Q}^\nu_{\;\mu\nu}=0$. Using \eqref{dector} and \eqref{nonmetricityECWfull} in 
\eqref{distortioncompl}, the distortion becomes
\begin{equation}\label{distcontECWfull}
\begin{split}
N_{\lambda \mu \nu} & =  \frac{1}{2} \left( \breve{T}_{\nu \lambda \mu} - \breve{T}_{\lambda \nu \mu} - \breve{T}_{\mu \nu \lambda} \right)+ \frac{1}{d-1} \left( g_{\mu \nu}T_\lambda - g_{\mu \lambda} T_\nu \right) \\
& + \Theta_\lambda g_{\mu \nu} - \Theta_\mu g_{\lambda \nu} - \Theta_\nu g_{\lambda \mu}+ \frac{1}{2} \left( \breve{Q}_{\lambda \mu \nu} + \breve{Q}_{\lambda \nu \mu} - \breve{Q}_{\mu \lambda \nu} \right) \\
& = \breve{K}_{\nu \lambda \mu} + \frac{1}{d-1} \left( g_{\mu \nu}T_\lambda - g_{\mu \lambda} T_\nu \right) + \Theta_\lambda g_{\mu \nu} - \Theta_\mu g_{\lambda \nu} - \Theta_\nu g_{\lambda \mu} + \breve{M}_{\lambda \mu \nu} \\
& = K_{\nu \lambda \mu} + M_{\lambda \mu \nu} ,
\end{split}
\end{equation}
where we defined the so-called disformation (also known as deflection tensor)
\begin{equation}\label{M}
\begin{split}
M_{\lambda \mu \nu} & = \frac{1}{2} \left( Q_{\lambda \mu \nu} + Q_{\lambda \nu \mu} - Q_{\mu \lambda \nu} \right) \\
& = \Theta_\lambda g_{\mu \nu} - \Theta_\mu g_{\lambda \nu} - \Theta_\nu g_{\lambda \mu}+ \frac{1}{2} \left( \breve{Q}_{\lambda \mu \nu} + \breve{Q}_{\lambda \nu \mu} - \breve{Q}_{\mu \lambda \nu} \right) \\
& =\Theta_\lambda g_{\mu \nu} - \Theta_\mu g_{\lambda \nu} - \Theta_\nu g_{\lambda \mu} + \breve{M}_{\lambda \mu \nu} ,
\end{split}
\end{equation}
which is symmetric in the last two indices. $\breve{K}_{\nu\lambda\mu}$ and $\breve{M}_{\nu\lambda\mu}$ 
are respectively the traceless part of $K_{\nu\lambda\mu}$ and $M_{\nu\lambda\mu}$,
\begin{equation}
\breve{K}_{\nu\lambda\mu} = \frac12\left(\breve{T}_{\nu\lambda\mu} - \breve{T}_{\lambda\nu\mu} - 
\breve{T}_{\mu\nu\lambda}\right), \qquad
\breve{M}_{\nu\lambda\mu} = \frac12\left(\breve{Q}_{\lambda\mu\nu} + \breve{Q}_{\lambda\nu\mu} - 
\breve{Q}_{\mu\lambda\nu}\right) . \label{MMt}
\end{equation}
From \eqref{connectionECW} one obtains for the connection
\begin{equation}
\begin{split}
\hat{\Gamma}^\lambda_{\; \mu \nu} & = \tilde{\Gamma}^\lambda_{\; \mu \nu} + \frac{1}{2} \left( \breve{T}^{\; \lambda}_{\nu\;\;  \mu} - \breve{T}^{\lambda}_{\; \nu \mu} - \breve{T}_{\mu \nu}^{\;\; \lambda} \right)+ \frac{1}{d-1} \left( g_{\mu \nu}T^\lambda - \delta_{\mu}^{\; \lambda} T_\nu \right) \\
& + \Theta^\lambda g_{\mu \nu} - \Theta_\mu \delta^\lambda_{\; \nu} - \Theta_\nu \delta^\lambda_{\; \mu} + \frac{1}{2} \left( \breve{Q}^\lambda _{ \; \mu \nu} + \breve{Q}^\lambda_{\; \nu \mu} - \breve{Q}^{\; \; \lambda }_{\mu \;\; \nu} \right) .
\end{split}
\end{equation}

{The explicit expression for the Ricci tensor $\hat{R}_{\rho \nu}$ of $\hat\nabla$ is given in the appendix (see \eqref{riccitensECWfull}), and it contains extra contributions from the traceless tensor $\breve{Q}_{\lambda\mu\nu}$.}
The homothetic curvature is still given by~\eqref{homot}, while the Ricci scalar is
\begin{equation}\label{ricciscalarECWfull}
\begin{split}
\hat{R} & = \tilde{R}  + \frac{(d-2)(1-d)}{(d-1)^2}T_\mu T^\mu + 2 \nabla_\mu T^\mu \\
& + (d-2)(1-d) \Theta_\mu \Theta^\mu + 2 (d-1) \nabla_\mu \Theta^\mu + 2 (2-d) \Theta^\mu T_\mu \\
& + \frac{1}{4} \left( \breve{T}_{\mu \nu \rho} \breve{T}^{\mu \nu \rho} - 2 \breve{T}_{\mu \nu \rho} \breve{Q}^{\mu \nu \rho}  + \breve{Q}_{\mu \nu \rho} \breve{Q}^{\mu \nu \rho}  \right) - \frac{1}{2} \left( \breve{T}_{\nu \rho \mu} \breve{T}^{\mu \nu \rho} + \breve{T}_{\mu \nu \rho} \breve{Q}^{\mu \nu \rho} + \breve{Q}_{\nu \rho \mu} \breve{Q}^{\mu \nu \rho} \right) .
\end{split}
\end{equation}
Observe that, by defining
\begin{equation}
\check{T}_\mu\equiv T_\mu + (d-1)\Theta_\mu , \qquad
\check{T}_{\mu\nu\rho}\equiv\breve{T}_{\mu\nu\rho} - \breve{Q}_{\mu\nu\rho} , \label{tcheckfullTT}
\end{equation}
where $\check{T}^\nu_{\;\mu \nu}=0$, and using the fact that the symmetries of
$\breve{T}_{\mu\nu\rho}$ and $\breve{Q}_{\mu\nu\rho}$ imply
\begin{equation}
\breve{T}_{\nu\rho\mu}\breve{Q}^{\mu\nu\rho} = 0 , \qquad
\breve{T}^{\mu\nu\rho}\breve{Q}_{\nu\rho\mu} = \breve{T}^{\mu\nu\rho}\breve{Q}_{\mu\rho\nu} ,
\end{equation}
one can shew that the Ricci scalar \eqref{ricciscalarECWfull} can be written as
\begin{equation}\label{tcheckeqfull}
\hat{R} = \tilde{R}  + \frac{(d-2)(1-d)}{(d-1)^2}\check{T}_\mu \check{T}^\mu + 2 \nabla_\mu \check{T}^\mu + \frac{1}{4} \check{T}_{\mu \nu \rho} \check{T}^{\mu \nu \rho} - \frac{1}{2} \check{T}_{\nu \rho \mu} \check{T}^{\mu \nu \rho} ,
\end{equation}
which corresponds to the Ricci scalar of a metric connection with nonvanishing torsion, whose trace and
traceless parts are now respectively given by $\check{T}_\mu$ and $\check{T}_{\mu\nu\rho}$.
This is analogous to the case in which one does not include a traceless contribution for the nonmetricity,
cf.~eq.~\eqref{tcheckeq}.

As before, we define an Einstein-Cartan-Weyl space by eq.~\eqref{symmcondECW}, which becomes
in the present context
\begin{equation}\label{pdeECWfull}
\begin{split}
& \tilde{R}_{\rho \nu}  + \frac{1}{d-1} \left[ (d-2) \nabla_{(\nu} T_{\rho)} + (d-3) T^\mu \breve{T}_{(\rho \nu ) \mu} \right]  + \frac{1}{(d-1)^2} \left[  (d-2) T_{\nu} T_{\rho} \right] \\
& + \frac{1}{4} \breve{T}_{\rho}^{\;\; \mu \sigma} \breve{T}_{\nu \mu \sigma}  - \frac{1}{2} \breve{T}_{\mu \sigma (\rho} \breve{T}_{\nu)}^{\;\;\;\mu \sigma} - \nabla_\mu \breve{T}_{(\rho \nu)}^{\;\;\;\;\;\mu}  \\
& + (d-2) \Theta_{\nu} \Theta_{\rho} + (d-2) \nabla_{(\nu} \Theta_{\rho)} + \frac{2(d-2)}{d-1} \Theta_{(\nu} T_{\rho)} + (d-3) \Theta^\mu \breve{T}_{(\nu \rho) \mu} \\
& + \frac{2-d}{d-1} T^\mu \breve{Q}_{\mu ( \nu \rho)} + \frac{d-4}{2(d-1)} T_\mu \breve{Q}^{\;\;\mu}_{\rho \;\; \nu} - (d-2) \Theta^\mu \breve{Q}_{\mu ( \nu \rho)} + \frac{d-4}{2} \Theta_\mu \breve{Q}^{\;\;\mu}_{\rho \;\; \nu} \\
& - \frac{1}{4} \breve{Q}_{\mu \rho \sigma} \breve{Q}^{\mu \;\;\sigma}_{\;\;\nu} + 2 \nabla^\mu \breve{Q}_{\mu (\nu \rho)} - \frac{1}{2} \nabla_\mu \breve{Q}^{\;\;\mu}_{\nu \;\; \rho} + \frac{1}{2} \breve{T}_{\mu (\rho}^{\;\;\;\;\;\sigma} \breve{Q}^{\mu}_{\;\; \nu) \sigma}  + \frac{1}{2} \breve{T}^{\mu \;\; \sigma}_{\;\;(\rho} \breve{Q}_{\nu) \sigma \mu} - \breve{T}_{(\rho}^{\;\;\mu \sigma} \breve{Q}_{\nu) \mu \sigma}  \\
&  = \frac{1}{d} g_{\rho \nu}  \Bigg [ \tilde{R} + \frac{d-2}{d-1} \nabla_\mu T^\mu  + \frac{d-2}{(d-1)^2} T_\mu T^\mu  + (d-2)\nabla_\mu \Theta^\mu + (d-2) \Theta_\mu \Theta^\mu + \frac{2(d-2)}{d-1} \Theta^\mu T_\mu \\
&  + \frac{1}{4} \left( \breve{T}^{\mu \tau \sigma} \breve{T}_{\mu \tau \sigma} + \breve{Q}^{\mu \tau \sigma} \breve{Q}_{\mu \tau \sigma} \right) - \frac{1}{2} \left( \breve{T}^{\mu \tau \sigma} \breve{T}_{\tau \sigma \mu} + \breve{Q}^{\mu \tau \sigma} \breve{Q}_{\tau \sigma \mu}  \right) - \breve{T}^{\mu \tau \sigma} \breve{Q}_{\mu \tau \sigma} \Bigg ] ,
\end{split}
\end{equation}
which represents a system of nonlinear partial differential equations characterizing an Einstein-Cartan-Weyl
manifold with the most general form of torsion and nonmetricity.

{Finally, we can consider the transformations \eqref{transfECW}.} In particular, we have
\begin{equation}
\breve{Q}^\lambda_{\;\mu\nu} \mapsto \breve{Q}^\lambda_{\;\mu\nu} .
\end{equation}
For the curvature tensors one still has the transformation laws given in \eqref{transfRECW}, so
that the Einstein-Cartan-Weyl equations \eqref{symmcondECW} are again {invariant under extended conformal transformations for arbitrary parameter $\xi$.}

\section{Einstein-Cartan action and scale invariant gravity}\label{ECactionphi}

Let us consider the action
\begin{equation}\label{actionfirst}
S = \int d^dx\sqrt{-g}\phi^2\left( R - \kappa\phi^{\frac4{d-2}}\right) ,
\end{equation}
where $R$ is the Ricci scalar \eqref{ricciscalar} of a torsionful but metric connection, $\phi$ denotes a
scalar field, and $\kappa$ is a constant. Along the same lines of \cite{Moon:2009zq}, \eqref{actionfirst}
can be rewritten as
\begin{equation}\label{actionEC}
S = \int d^dx \sqrt{-g} \phi^2 \left( \tilde{R}  - \frac{d-2}{d-1}T_\mu T^\mu + 2 \nabla_\mu T^\mu + \frac{1}{4} \breve{T}_{\mu \nu \rho} \breve{T}^{\mu \nu \rho} - \frac{1}{2} \breve{T}_{\nu \rho \mu} \breve{T}^{\mu \nu \rho} - \kappa \phi^{\frac{4}{d-2}} \right),
\end{equation}
with $\tilde{R}$ the scalar curvature of the Levi-Civita connection. One easily shows that \eqref{actionEC} is 
invariant under
\begin{equation}
g_{\mu\nu}\mapsto e^{2\omega} g_{\mu\nu}, \qquad
\phi\mapsto e^{\frac{2-d}2\omega}\phi, \qquad
T_\mu\mapsto T_\mu + (d-1)\partial_\mu\omega, \qquad
\breve{T}^\lambda_{\;\mu\nu}\mapsto\breve{T}^\lambda_{\;\mu\nu}. \label{tr2T}
\end{equation}
Using the traceless part of the contorsion defined in \eqref{MMt}, the action \eqref{actionEC} becomes
\begin{equation}\label{actionECK}
S = \int d^dx\sqrt{-g}\phi^2\left(\tilde{R} - \frac{d-2}{d-1}T_\mu T^\mu + 2\nabla_\mu T^\mu - 
\breve{K}_{\nu\rho\mu}\breve{K}^{\mu\nu\rho} - \kappa\phi^{\frac4{d-2}}\right),
\end{equation}
and its variation w.r.t.~$T_\mu$ and $\breve{K}_{\nu\rho\mu}$ yields respectively
\begin{equation}
T_\mu = -\frac{2(d-1)}{d-2}\frac{\nabla_\mu\phi}{\phi}, \qquad
\breve{K}_{\mu [\nu\rho]} = 0. \label{eqKmurhonu}
\end{equation}
Notice that $T_\mu$ can be eliminated by {an extended conformal transformation} and is thus pure gauge.
Using the definition \eqref{MMt} and the fact that the traceless part of the torsion is antisymmetric
in the last two indices, we get $\breve{T}_{\mu\nu\rho} =2\breve{K}_{\mu [\nu\rho]}=0$,
and therefore also $\breve{K}_{\mu\nu\rho} =0$, in agreement
with \cite{HelayelNeto:1999tm, Moon:2009zq}.

Varying the action \eqref{actionECK} w.r.t.~$g_{\mu\nu}$ and $\phi$ leads to
\begin{subequations}
\begin{align}
& \phi^2 \left( \tilde{R}_{\mu \nu} - \frac{1}{2} g_{\mu \nu} \tilde{R} \right) + \frac{2 d}{d-2} \nabla_\mu \phi \nabla_\nu \phi - 2 \phi \nabla_\nu \nabla_\mu \phi \nonumber \\ 
& + 2 g_{\mu \nu} \phi \nabla_\rho \nabla^\rho \phi - \frac{2}{d-2}  g_{\mu \nu} \nabla_\rho \phi \nabla^\rho \phi + \frac{1 }{2} g_{\mu \nu} \kappa \phi^{\frac{2d}{d-2}} =0 , \label{eom1EC} \\
& \phi \tilde{R} - \frac{4(d-1)}{d-2}\nabla_\rho \nabla^\rho \phi - \frac{d}{d-2} \kappa \phi^{\frac{d+2}{d-2}} =0, \label{eom2EC}
\end{align}
\end{subequations}
where we have used the expression for $T_\nu$ in \eqref{eqKmurhonu} as well as
$\breve{K}_{\mu\nu\rho} =0$. Observe that the trace of \eqref{eom1EC} implies \eqref{eom2EC}, which
can be understood as a consequence of $\phi$ being pure gauge.

Let us now consider the {action}
\begin{equation}\label{actionECnew}
S = \int d^dx \sqrt{-g} \left[ \phi^2  \tilde{R} + \frac{4(d-1)}{d-2} \nabla_\mu \phi \nabla^\mu \phi - \kappa \phi^{\frac{2 d}{d-2}} \right] ,
\end{equation}
which is called scale invariant {(also known as conformal gravity)}. It turns out that the equations of motion following
from \eqref{actionECnew} are precisely \eqref{eom1EC} and \eqref{eom2EC} obtained from
\eqref{actionECK} after having used the expressions for the torsion. The actions \eqref{actionfirst} and 
\eqref{actionECnew} describe thus the same dynamics. Notice also that, plugging $T_\mu$
(cf.~\eqref{eqKmurhonu}) and $\breve{K}_{\mu\nu\rho} =0$ into \eqref{actionECK}, one gets, up to a
surface term\footnote{The surface term is $\int d^dx\sqrt{-g}\left[-\frac{4(d-1)}{d-2}\nabla_\mu\left(\phi 
\nabla^\mu\phi\right)\right]$.}, the conformal gravity action \eqref{actionECnew} (see
also \cite{Moon:2009zq}).

One can also show that the action \eqref{actionfirst} implies that the spacetime is Einstein with torsion, which is a completely new result. 
To see this, observe that eq.~\eqref{eom1EC} can be rewritten as
\begin{equation}\label{eom1newnewR}
\tilde{R}_{\mu \nu} + \frac{2 d}{d-2} \frac{\nabla_\mu \phi \nabla_\nu \phi}{\phi^2} - 2 \frac{\nabla_\nu \nabla_\mu \phi}{\phi} = \frac{1}{d} g_{\mu \nu} \left( \frac{d}{2} \tilde{R} - 2d \frac{\nabla_\rho \nabla^\rho \phi}{\phi} + \frac{2d}{d-2} \frac{\nabla_\rho \phi \nabla^\rho \phi}{\phi^2} - \frac{d}{2} \kappa \phi^{\frac{4}{d-2}} \right) .
\end{equation}
Using also \eqref{eom2EC}, this can be cast into the form
\begin{equation}\label{eom1newnew}
\tilde{R}_{\mu \nu} + \frac{2 d}{d-2} \frac{\nabla_{\mu} \phi \nabla_{\nu} \phi}{\phi^2} - 2 \frac{\nabla_{\mu} \nabla_{\nu} \phi}{\phi} = \frac{1}{d} g_{\mu \nu} \left(\tilde{R}- 2\frac{\nabla_\rho \nabla^\rho \phi}{\phi} + \frac{2d}{d-2} \frac{\nabla_\rho \phi \nabla^\rho \phi}{\phi^2}\right).
\end{equation}
On the other hand, consider the system \eqref{pdeEC} characterizing an Einstein-Cartan manifold,
and use the result \eqref{eqKmurhonu} for the trace part of the torsion as well as
$\breve{T}_{\mu\nu\rho}=0$. Then \eqref{pdeEC} boils down precisely to \eqref{eom1newnew}.

Let us also observe that, as already mentioned in \cite{Moon:2009zq}, conformal (Weyl) invariance allows to rescale $\phi\mapsto e^{\frac{2-d}{d}\omega}\phi$. One can use this freedom to gauge fix
$\phi=1/(4\sqrt{\pi G})$, where $G$ is Newton's constant. Then the action \eqref{actionECnew} becomes
\begin{equation}\label{actioncosmconst}
S = \frac1{16\pi G}\int d^d x\sqrt{-g}\left(\tilde{R} - 2\Lambda\right) ,
\end{equation}
where we chose $\kappa = 2\Lambda(16\pi G)^{2/(d-2)}$. The Einstein-Hilbert action with cosmological 
constant can thus be viewed as a gauge fixed version of the {action} \eqref{actionECnew}.

Finally, let us recall that the trace part of the torsion can also be interpreted as the trace part of the
nonmetricity (cf.~sec.~\ref{subsec:compEW}). If we set the traceless part of the torsion to zero, this
leads to the action
\begin{equation}\label{actionEWW}
S = \int d^dx \sqrt{-g} \phi^2 \left( W - \kappa \phi^{\frac{4}{d-2}} \right) ,
\end{equation}
which is invariant under
\begin{equation}
g_{\mu\nu}\mapsto e^{2\omega} g_{\mu\nu} , \qquad
\phi\mapsto e^{\frac{2-d}2\omega}\phi , \qquad
\Theta_\mu\mapsto\Theta_\mu + \partial_\mu \omega .
\end{equation}
The variation of \eqref{actionEWW} w.r.t.~$\Theta_\mu$ yields
\begin{equation}
\Theta_\mu = -\frac2{d-2}\frac{\nabla_\mu\phi}{\phi} . \label{eqThetamu} 
\end{equation}
Again, one can easily show that the actions \eqref{actionEWW} and \eqref{actionECnew} describe the same 
dynamics.
\eqref{actionEWW} implies that the spacetime is Einstein-Weyl, where the Weyl vector is given by \eqref{eqThetamu}, and is thus pure gauge. Notice in this context that there is no known action principle
that leads to the Einstein-Weyl equations with non-exact Weyl vector.

\section{Einstein-Hilbert action coupled to a 3-form as Einstein-Cartan gravity}\label{ECactionH}

The Einstein-Hilbert action coupled to a 3-form field strength reads
\begin{equation}\label{Haction}
S_1 = \int d^d x \sqrt{-g}  \left( \tilde{R} - \frac{1}{12} H_{\mu \nu \rho} H^{\mu \nu \rho}  \right) ,
\end{equation}
where $H_{\mu\nu\rho}$ is given in terms of a gauge potential $B_{\mu\nu}$,
\begin{equation}\label{defH}
H_{\mu\nu\rho} =\partial_\mu B_{\nu\rho} + \partial_\nu B_{\rho\mu} + \partial_\rho B_{\mu\nu}, \qquad 
B_{\mu\nu} = - B_{\nu\mu} .
\end{equation}
The variation of \eqref{Haction} w.r.t.~$B_{\mu\nu}$ leads to
\begin{equation}\label{eqH}
\nabla^\mu H_{\mu \nu \rho} = 0,
\end{equation}
while varying $g^{\rho\nu}$ gives
\begin{equation}\label{eomgmunuHH}
\tilde{R}_{\rho\nu} - \frac12 g_{\rho\nu}\tilde{R} + \frac1{24} g_{\rho\nu} H_{\mu\tau\sigma}
H^{\mu\tau\sigma} - \frac14 H_{\rho}^{\; \;\mu\sigma} H_{\nu\mu\sigma} = 0.
\end{equation}
On the other hand, consider the system \eqref{pdeEC} satisfied by an Einstein manifold with torsion.
Assume that $T_\mu=0$ and take $\breve{T}_{\mu\nu\rho}$ to be completely antisymmetric. Then
\eqref{pdeEC} boils down to
\begin{equation}\label{pdeEC-H}
\tilde{R}_{\rho\nu} - \frac14\breve{T}_{\mu\sigma\nu}\breve{T}_\rho^{\;\;\mu\sigma} = 
\frac1d g_{\rho\nu}\left(\tilde{R} - \frac14\breve{T}^{\mu\tau\sigma}\breve{T}_{\mu\tau\sigma}\right).
\end{equation}
We would like to compare this with \eqref{eomgmunuHH}. To this end, take the trace of
\eqref{eomgmunuHH}, which leads to
\begin{equation}\label{traceRH}
\tilde{R} = \frac{d-6}{12(d-2)}H^2, \qquad H^2\equiv H_{\mu\tau\sigma} H^{\mu\tau\sigma}.
\end{equation}
Now subtract its trace part from \eqref{eomgmunuHH} to obtain
\begin{equation}
\tilde{R}_{\rho\nu} - \frac1d g_{\rho\nu}\tilde{R} - \frac14 H_\rho^{\;\;\mu\sigma} H_{\nu\mu\sigma}
+ \frac1{4d} g_{\rho\nu} H^2 = 0,
\end{equation}
which coincides precisely with \eqref{pdeEC-H} if we identify $H_{\mu\nu\rho}=\breve{T}_{\mu\nu\rho}$.
The equations of motion following from \eqref{Haction} can thus be interpreted as implying that the spacetime is Einstein with
skew-symmetric torsion $H_{\mu\nu\rho}$ satisfying \eqref{eqH}. Notice however that the equations
\eqref{eomgmunuHH} are more restrictive than \eqref{pdeEC-H}, since they contain in addition the trace
part \eqref{traceRH}, while \eqref{pdeEC-H} is traceless. This is somehow reminiscent of hyper Cauchy-Riemann (hyper-CR, or Gauduchon-Tod) spaces \cite{Gauduchon:1998}, where on top of the (trace-free) Einstein-Weyl equations
there is a constraint on the scalar curvature.

Quite remarkably, the equations \eqref{eqH}, \eqref{eomgmunuHH} can also be retrieved from the 
constrained action
\begin{eqnarray}
S_2 &=& \int d^d x\sqrt{-g} \left[ R + \lambda^{\mu\nu\rho}\left(\breve{T}_{\mu\nu\rho} - \frac1{\sqrt3}
\left(\partial_\mu B_{\nu\rho} + \partial_\nu B_{\rho\mu} + \partial_\rho B_{\mu\nu}\right)\right) \right]
\nonumber \\
&=& \int d^d x \sqrt{-g} \left[ \tilde{R} - \frac{d-2}{d-1} T_\mu T^\mu + 2\nabla_\mu T^\mu + 
\frac14\breve{T}_{\mu\nu\rho}\breve{T}^{\mu\nu\rho} -\frac12\breve{T}_{\nu\rho\mu}
\breve{T}^{\mu\nu\rho}\right. \nonumber \\
&&\qquad\qquad\quad\left. + \lambda^{\mu\nu\rho}\left(\breve{T}_{\mu\nu\rho} - \frac1{\sqrt3}
(\partial_\mu B_{\nu\rho} + \partial_\nu B_{\rho\mu} + \partial_\rho B_{\mu\nu})\right) \right],
\label{actionEConlyR}
\end{eqnarray}
where $R$ denotes the scalar curvature of a torsionful but metric connection (cf.~\eqref{ricciscalar}),
$\lambda^{\mu\nu\rho}$ is a Lagrange multiplier, and $B_{\mu\nu}$ is antisymmetric.
The variation of \eqref{actionEConlyR} w.r.t. $T_\mu$, $B_{\mu\nu}$, $\lambda^{\mu\nu\rho}$, 
$\breve{T}_{\mu\nu\rho}$ and $g^{\mu\nu}$ gives respectively
\begin{equation}\label{Tmu=0ineom}
T_\mu =0, \qquad \nabla_\mu\lambda^{[\mu\nu\rho]} = 0,
\end{equation}
\begin{equation}
\breve{T}_{\mu\nu\rho} = \frac1{\sqrt3}(\partial_\mu B_{\nu\rho} + \partial_\nu B_{\rho\mu} + 
\partial_\rho B_{\mu\nu}), \label{T=dB}
\end{equation}
\begin{equation}
\lambda^{\mu\nu\rho} = \frac12\left(\breve{T}^{\nu\rho\mu} + \breve{T}^{\rho\mu\nu} -
\breve{T}^{\mu\nu\rho}\right), \label{lagrange}
\end{equation}
\begin{equation}\label{eomgwithTTproptoH}
\tilde{R}_{\mu \nu} - \frac{1}{2} g_{\mu \nu} \tilde{R}+ \frac{1}{8}g_{\mu \nu} \breve{T}_{\tau \rho \sigma} \breve{T}^{\tau \rho \sigma}- \frac{3}{4} \breve{T}_{\mu}^{\; \; \tau \rho} \breve{T}_{\nu \tau \rho} =0,
\end{equation}
where we already used $T_\mu=0$ in \eqref{eomgwithTTproptoH}. \eqref{T=dB}
implies that the traceless part of the torsion is completely antisymmetric, and thus \eqref{lagrange}
reduces to
\begin{equation}
\lambda^{\mu\nu\rho} = \frac12\breve{T}^{\mu\nu\rho}.
\end{equation}
Plugging this into the last eq.~of \eqref{Tmu=0ineom} leads to
\begin{equation}
\nabla_\mu\breve{T}^{\mu\nu\rho} = 0. \label{nablaT=0}
\end{equation}
Finally, using \eqref{T=dB} in \eqref{nablaT=0} and \eqref{eomgwithTTproptoH}, one gets
precisely \eqref{eqH} and \eqref{eomgmunuHH}. The actions $S_1$ and $S_2$ describe therefore the
same dynamics.

\section{Discussion}\label{Discussion}

Motivated by the interest in connections with torsion and nonmetricity both from the physical and the 
mathematical point of view, we {first} generalized here some results that appeared previously in the literature. 
In particular, we considered Einstein spaces with nonvanishing torsion that has both a trace and a traceless
part (Einstein-Cartan manifolds), and showed that the resulting field equations are {invariant under extended conformal transformations.}
We then compared our results to Einstein manifolds with zero torsion but nonvanishing nonmetricity,
where the latter is given in terms of the Weyl vector $\Theta_\mu$ (Einstein-Weyl spaces).
We saw that, if the traceless part of the torsion is set to zero, then the system of partial differential
equations characterizing Einstein-Cartan spaces exactly coincides with the Einstein-Weyl equations
if the torsion trace $T_\mu$ is replaced by $(d-1)\Theta_\mu$.
Subsequently, we extended our analysis to the case of Einstein manifolds with both torsion and
nonmetricity (Einstein-Cartan-Weyl spaces), allowing for both a trace and a traceless part of the
nonmetricity tensor.

Moreover, we considered actions involving scalar curvatures obtained from torsionful
or nonmetric connections, and investigated their relations with other gravitational theories{, obtaining completely new results in this context.} In particular,
we analyzed a {conformally (Weyl) invariant} action with torsion and its relation with scale invariant gravity, which involves a scalar $\phi$, and found that they reproduce the same dynamics. {Furthermore, we have shown that the action \eqref{actionfirst} implies that the spacetime is Einstein with torsion.} 
Then, the Einstein-Hilbert action coupled to a three-form field strength $H_{\mu\nu\rho}$ was
considered, and it was shown that its equations of motion imply that the manifold is Einstein with
skew-symmetric torsion. Furthermore, it turned out that the equations of motion of Einstein gravity
coupled to a three-form may also be retrieved from a constrained action that contains the scalar
curvature of a connection with torsion.
Let us stress that in this paper we concentrated on the vacuum, without considering the presence of matter.

{Among the solutions to Einstein's field equations, Einstein spaces are of particular relevance in physics, think for instance of the Kerr-(A)dS solution or of string compactifications on e.g. Sasaki-Einstein manifolds. Since nature could accommodate for torsion and nonmetricity, it seems reasonable to generalize the concept of Einstein spaces to torsionful and nonmetric connections.}

The manifolds analyzed in this paper may {also} have applications in the classification {and physical study} of (fake) supersymmetric supergravity solutions in the same way as Einstein-Weyl manifolds provide the base space for fake supersymmetric solutions in de Sitter supergravity \cite{Meessen:2009ma, Gutowski:2009vb, Grover:2009ms, Gutowski:2011gc, Dunajski:2016rtx, Randall:2017zlm}.
{Under the physical point of view, this analysis is particularly relevant in higher dimensions, since, in $d>4$, it is highly nontrivial to determine whether a given near-horizon geometry can be extended to a full black hole solution (due to the fact that the strong uniqueness theorems
that hold in four dimensions \cite{Hawking:1971vc,Israel:1967wq,Carter:1971zc,Robinson:1975bv, Israel:1967za,Mazur:1982db} break down and there exist different black holes with the same asymptotic 
charges and different black hole solutions with the same near-horizon geometry). Progress in classifying near-horizon geometries can help to face this problem, as it was proven in \cite{Dunajski:2016rtx}, where the authors, after having showed that a class of solutions of minimal supergravity in five dimensions is given by lifts of three-dimensional Einstein-Weyl structures of hyper-CR type, considered the task of reconstructing all supersymmetric solutions from such near-horizon geometry, demonstrating that the moduli space of infinitesimal supersymmetric transverse deformations of the near-horizon data is finite-dimensional if the spatial section of the horizon is compact.}

{Always in this context, a new result has recently been obtained} in \cite{Klemm:2019izb}, where it has been shown that the horizon geometry for supersymmetric black hole solutions of minimal five-dimensional gauged supergravity is that of a particular Einstein-Cartan-Weyl structure in three dimensions, involving the trace and traceless part of both torsion and nonmetricity, and obeying some precise constraint; in the limit of zero cosmological constant, the set of nonlinear partial differential equations characterizing this Einstein-Cartan-Weyl structure reduces to that of a hyper-CR Einstein-Weyl structure in the Gauduchon gauge, which was shown in \cite{Dunajski:2016rtx} to be the horizon geometry in the ungauged BPS (Bogomol'nyi-Prasad-Sommefield) case.

The analysis of this paper might also be extended in other directions. In particular, it would be interesting
to generalize the construction of \cite{Cacciatori:2005wz} concerning the Chern-Simons
formulation of three-dimensional gravity involving torsion and nonmetricity, and the recent
results presented in \cite{Penas:2018mnc} in the context of double field theory. One could also investigate 
possible generalizations of \cite{Boulanger:2006tg, Baekler:2006vw}.

On the other hand, a future development of our work may consist in  possible generalizations of the
Jones-Tod correspondence \cite{Jones:1985pla} between selfdual conformal four-manifolds with a
conformal vector field and abelian monopoles on Einstein-Weyl spaces in three dimensions.
Especially one could ask whether Einstein-Cartan-Weyl manifolds can arise in a similar way by
symmetry reduction from higher dimensions.

Finally, a further direction for future research would be a geometrical investigation of the results on 
unconventional supersymmetry presented recently in \cite{Andrianopoli:2018ymh}, where torsion plays
a fundamental role, under the perspective developed here.

\appendix

\section{Riemann and Ricci tensors}

The Riemann tensor of the Einstein-Cartan connection $\Gamma^\lambda_{\; \mu\nu}$ introduced in section \ref{ECmani} reads
\begin{equation}\label{riemanntors}
\begin{split}
R^\lambda_{\;\rho\mu\nu} & = \partial_\mu\Gamma^\lambda_{\;\nu\rho} - \partial_\nu 
\Gamma^\lambda_{\;\mu\rho} + \Gamma^\lambda_{\;\mu\sigma}\Gamma^\sigma_{\;\nu\rho} - 
\Gamma^\lambda_{\;\nu\sigma}\Gamma^\sigma_{\;\mu\rho} \\
& = \tilde{R}^\lambda_{\; \rho \mu \nu} + \frac{1}{d-1} \left[g_{\rho \nu} \nabla_\mu T^\lambda - g_{\rho \mu} \nabla_\nu T^\lambda + \delta_\mu^{\; \lambda} \nabla_\nu T_{\rho} -  \delta_\nu^{\; \lambda} \nabla_\mu T_{\rho}  \right] \\
& +  \frac{1}{d-1} \left[ \frac{1}{2} \delta_\nu^{\;\lambda} T^\sigma \left( - \breve{T}_{\sigma \mu \rho} + \breve{T}_{\mu \rho \sigma} - \breve{T}_{\rho \mu \sigma} \right) + \frac{1}{2} \delta_\mu^{\; \lambda} T^\sigma \left( - \breve{T}_{\sigma \nu \rho} + \breve{T}_{\nu \rho \sigma} + \breve{T}_{\rho \nu \sigma} \right)  \right] \\
& + \frac{1}{d-1} \left[ T^\lambda \breve{T}_{\rho \mu \nu}  - T_\rho\breve{T}^\lambda_{\;\mu \nu}  \right] \\ 
& + \frac{1}{d-1} \Bigg \lbrace  \frac{1}{2} T^\sigma \left[  g_{\nu \rho} \left(  \breve{T}_{\sigma \;\;\;\mu}^{\;\;\lambda} + \breve{T}^\lambda_{\; \mu \sigma} + \breve{T}_{\mu\;\;\;\sigma}^{\;\;\lambda}  \right) - g_{\mu \rho} \left(  \breve{T}_{\sigma \;\;\;\nu}^{\;\;\lambda} + \breve{T}^\lambda_{\; \nu \sigma} + \breve{T}_{\nu\;\;\;\sigma}^{\;\;\lambda} \right) \right] \Bigg \rbrace \\
& + \frac{1}{(d-1)^2} \left[ g_{\nu \rho} T_\mu T^\lambda - g_{\mu \rho} T_\nu T^\lambda + \left( g_{\rho \mu} \delta_\nu^{\; \lambda} - g_{\rho \nu} \delta_\mu^{\; \lambda} \right) T_\sigma T^\sigma + T_\rho \left(\delta_\mu^{\; \lambda} T_\nu - \delta_\nu^{\; \lambda} T_\mu \right) \right] \\
& + \frac{1}{4} \left[ \breve{T}^{\lambda \;\; \sigma}_{\;\; \nu}  \breve{T}_{\mu \rho \sigma} + \breve{T}_{\mu \rho}^{\;\;\;\sigma} \left( \breve{T}^{\; \lambda}_{\sigma \;\; \nu} + \breve{T}^{\; \lambda}_{\nu \;\; \sigma} \right)  - \breve{T}_{\sigma \mu \rho} \left( \breve{T}^{\sigma \lambda}_{\;\;\;\; \nu} + \breve{T}^{\lambda \;\; \sigma}_{\;\; \nu} + \breve{T}_{\nu}^{\; \lambda \sigma} \right) + \left( \breve{T}^{\lambda \;\;\sigma}_{\;\;\nu} + \breve{T}_{\nu}^{\;\lambda \sigma} \right) \breve{T}_{\rho \mu \sigma}       \right] \\
& + \frac{1}{4} \left[ \breve{T}_{\sigma \;\;\; \nu}^{\;\;\lambda} \breve{T}_{\rho \mu}^{\;\;\; \sigma} - \breve{T}_{\rho \;\;\; \mu}^{\;\; \lambda} \left(\breve{T}_{\nu \rho}^{\;\;\;\sigma} + \breve{T}_{\rho \nu}^{\;\;\; \sigma} \right) + \breve{T}_{\sigma \nu \rho} \left( \breve{T}^{\sigma \lambda}_{\;\;\;\; \mu} + \breve{T}^{\lambda \;\;\; \sigma}_{\;\;\; \mu}+ \breve{T}_\mu^{\;\;\lambda \sigma} \right) \right] \\
& - \frac{1}{4} \left[ \left(\breve{T}^{\lambda \;\;\; \sigma}_{\;\;\;\mu}+ \breve{T}_\mu^{\;\;\lambda \sigma} \right) \left( \breve{T}_{\nu \rho \sigma} + \breve{T}_{\rho \nu \sigma}  \right)    \right] \\
& + \frac{1}{2} \left[\nabla_\mu \breve{T}^\lambda_{\; \nu \rho} + \nabla_\mu \breve{T}_{\nu \;\;\;\rho}^{\;\; \lambda}+ \nabla_\mu \breve{T}_{\rho \;\;\;\nu}^{\;\;\lambda} - \nabla_\nu \breve{T}^\lambda_{\;\mu \rho} - \nabla_\nu \breve{T}_{\mu \;\;\; \rho}^{\;\; \lambda} - \nabla_\nu \breve{T}_{\rho \;\;\; \mu}^{\;\; \lambda}\right] ,
\end{split}
\end{equation}
where $\tilde{R}^\lambda_{\;\rho\mu\nu}$ and $\nabla$ denote respectively the Riemann tensor and
the covariant derivative of the Levi-Civita connection. The first line of \eqref{riemanntors} follows from the definition $\left[D_\mu,D_\nu\right]\omega_\rho + T^\sigma_{\; \mu\nu} D_\sigma\omega_\rho =
-R^\lambda_{\; \rho\mu\nu}\omega_\lambda$, where $D$ denotes the connection with coefficients
$\Gamma$.
The corresponding Ricci tensor is given by
\begin{equation}\label{riccitens}
\begin{split}
R_{\rho \nu} & = R^\mu_{\; \rho \mu \nu} = \tilde{R}_{\rho \nu} +  \frac{1}{d-1} \left[ g_{\rho \nu} \nabla_\mu T^\mu + (d-2) \nabla_\nu T_\rho  \right] + \frac{1}{(d-1)^2} \left[ (2-d) g_{\nu \rho} T_\mu T^\mu + (d-2) T_\nu T_\rho \right]  \\
& + \frac{1}{d-1} \Bigg \lbrace \frac{1}{2}T^\mu \left[ (2-d) \left(\breve{T}_{\mu \nu \rho} - \breve{T}_{\nu \rho \mu} \right) + (d-4) \breve{T}_{\rho \nu \mu} \right]  \Bigg \rbrace \\
& + \frac{1}{4} \breve{T}_\nu^{\;\mu \sigma} \breve{T}_{\rho \mu \sigma} + \frac{1}{2} \left( \breve{T}_{\mu \nu \sigma} \breve{T}_\rho^{\; \mu \sigma} + \nabla_\mu \breve{T}^\mu_{\;\nu \rho} - \nabla_\mu \breve{T}_{\nu \rho}^{\;\;\;\;\mu} - \nabla_\mu \breve{T}_{\rho \nu}^{\;\;\;\; \mu} \right) .
\end{split}
\end{equation}

On the other hand, the Ricci tensor of the Einstein-Cartan-Weyl connection $\hat{\Gamma}^\lambda_{\;\mu\nu}$ introduced in section \ref{ECWmani} is
\begin{equation}\label{riccitensECW}
\begin{split}
\hat{R}_{\rho \nu} & = \hat{R}^\mu_{\; \rho \mu \nu} = \tilde{R}_{\rho \nu} +  \frac{1}{d-1} \left[ g_{\rho \nu} \nabla_\mu T^\mu + (d-2) \nabla_\nu T_\rho  \right] + \frac{1}{(d-1)^2} \left[ (2-d) g_{\nu \rho} T_\mu T^\mu + (d-2) T_\nu T_\rho \right]  \\
& + \frac{1}{d-1} \Bigg \lbrace \frac{1}{2}T^\mu \left[ (2-d) \left(\breve{T}_{\mu \nu \rho} - \breve{T}_{\nu \rho \mu} \right) + (d-4) \breve{T}_{\rho \nu \mu} \right]  \Bigg \rbrace \\
& + \frac{1}{4} \breve{T}_\nu^{\;\mu \sigma} \breve{T}_{\rho \mu \sigma} + \frac{1}{2} \left( \breve{T}_{\mu \nu \sigma} \breve{T}_\rho^{\; \mu \sigma} + \nabla_\mu \breve{T}^\mu_{\;\nu \rho} - \nabla_\mu \breve{T}_{\nu \rho}^{\;\;\;\;\mu} - \nabla_\mu \breve{T}_{\rho \nu}^{\;\;\;\; \mu} \right) \\
& + (d-2) \Theta_\nu \Theta_\rho + (2-d) \Theta_\mu \Theta^\mu g_{\nu \rho} + g_{\nu \rho} \nabla_\mu \Theta^\mu + (d-1) \nabla_\nu \Theta_\rho - \nabla_\rho \Theta_\nu \\
& + \frac{1}{d-1} \left[ (d-2)\Theta_\rho T_\nu + (d-2) \Theta_\nu T_\rho + 2(2-d) g_{\nu \rho} \Theta^\mu T_\mu  \right] \\
& + \frac{d-2}{2} \Theta^\mu \left(\breve{T}_{\mu \nu \rho} - \breve{T}_{\nu \rho \mu} \right) + \frac{d-4}2\Theta^\mu\breve{T}_{\rho\nu\mu} ,
\end{split}
\end{equation}
where $\nabla$ denotes again the Levi-Civita connection.

Finally, adding a traceless part to the nonmetricity tensor, we have that the Ricci tensor of $\hat\nabla$ reads, explicitly,
\begin{equation}\label{riccitensECWfull}
\begin{split}
\hat{R}_{\rho \nu} & = \hat{R}^\mu_{\; \rho \mu \nu} = \tilde{R}_{\rho \nu} +  \frac{1}{d-1} \left[ g_{\rho \nu} \nabla_\mu T^\mu + (d-2) \nabla_\nu T_\rho  \right] + \frac{1}{(d-1)^2} \left[ (2-d) g_{\nu \rho} T_\mu T^\mu + (d-2) T_\nu T_\rho \right]  \\
& + \frac{1}{d-1} \Bigg \lbrace \frac{1}{2}T^\mu \left[ (2-d) \left(\breve{T}_{\mu \nu \rho} - \breve{T}_{\nu \rho \mu} \right) + (d-4) \breve{T}_{\rho \nu \mu} \right]  \Bigg \rbrace \\
& + \frac{1}{4} \breve{T}_\nu^{\;\mu \sigma} \breve{T}_{\rho \mu \sigma} + \frac{1}{2} \left( \breve{T}_{\mu \nu \sigma} \breve{T}_\rho^{\; \mu \sigma} + \nabla_\mu \breve{T}^\mu_{\;\nu \rho} - \nabla_\mu \breve{T}_{\nu \rho}^{\;\;\;\;\mu} - \nabla_\mu \breve{T}_{\rho \nu}^{\;\;\;\; \mu} \right) \\
& + (d-2) \Theta_\nu \Theta_\rho + (2-d) \Theta_\mu \Theta^\mu g_{\nu \rho} + g_{\nu \rho} \nabla_\mu \Theta^\mu + (d-1) \nabla_\nu \Theta_\rho - \nabla_\rho \Theta_\nu \\
& + \frac{1}{d-1} \left[ (d-2)\Theta_\rho T_\nu + (d-2) \Theta_\nu T_\rho + 2(2-d) g_{\nu \rho} \Theta^\mu T_\mu  \right] \\
& + \frac{d-2}{2} \Theta^\mu \left(\breve{T}_{\mu \nu \rho} - \breve{T}_{\nu \rho \mu} \right) + \frac{d-4}{2} \Theta^\mu \breve{T}_{\rho \nu \mu} \\
& + \frac{1}{d-1} \Bigg \lbrace \frac{1}{2}T^\mu \left[ (2-d) \left(\breve{Q}_{\mu \nu \rho} + \breve{Q}_{\nu \rho \mu} \right) + (d-4) \breve{Q}_{\rho \mu \nu} \right]  \Bigg \rbrace \\
& - \frac{d-2}{2} \Theta^\mu \left(\breve{Q}_{\mu \nu \rho} + \breve{Q}_{\nu \rho \mu} \right) + \frac{d-4}{2} \Theta^\mu \breve{Q}_{\rho \mu \nu} \\
& - \frac{1}{4} \breve{Q}_{\mu \rho \sigma} \breve{Q}^{\mu \;\; \sigma}_{\;\; \nu} + \frac{1}{2} \left( \breve{Q}_{\nu}^{\;\; \mu \sigma} \breve{Q}_{\rho \mu \sigma} - \breve{Q}_{\nu}^{\;\; \mu \sigma} \breve{Q}_{\rho \sigma \mu} - \nabla_\mu \breve{Q}^{\;\;\mu}_{\nu \;\; \rho} + \nabla_\mu \breve{Q}_{\nu \rho}^{\;\;\;\;\mu} + \nabla_\mu \breve{Q}_{\rho \nu}^{\;\;\;\; \mu} \right) \\
& + \frac{1}{2} \breve{T}^{\mu \;\; \sigma}_{\;\; \rho} \breve{Q}_{\mu \nu \sigma} - \frac{1}{2} \breve{T}_\rho^{\;\;\mu \sigma} \breve{Q}_{\nu \mu \sigma} - \frac{1}{2}\breve{T}_\nu^{\;\;\mu \sigma} \breve{Q}_{\rho \mu \sigma} + \frac{1}{2} \breve{T}^{\mu \;\; \sigma}_{\;\;\nu} \left( \breve{Q}_{\rho \sigma \mu} - \breve{Q}_{\rho \mu \sigma} \right) ,
\end{split}
\end{equation}
which, indeed, now contains extra contributions from the traceless tensor $\breve{Q}_{\lambda\mu\nu}$.

\section{Acknowledgements}

L.~R. acknowledges illuminating discussions with L.~Andrianopoli, B.~L.~Cerchiai, M.~Trigiante and
J.~Zanelli. D.~K.~is supported partly by INFN.

\end{document}